\documentclass[journal,twocolumn,twoside]{IEEEtran} 

\usepackage[T1]{fontenc}
\usepackage{CJKutf8}
\usepackage[english]{babel}

\usepackage{graphicx}
\usepackage[export]{adjustbox}

\graphicspath{{./Figs/}}
\usepackage{subfigure}
\usepackage{stfloats}
\ifCLASSINFOpdf
\else
\fi

\usepackage{amssymb}
\usepackage{amsmath}
\usepackage[cmintegrals]{newtxmath}
\usepackage{stackengine}

\usepackage{multirow}
\usepackage{booktabs}
\newcommand{\tabitem}{~~\llap{\textbullet}~~}

\usepackage{soul,xcolor} 
\setstcolor{red}

\usepackage{xpatch}
\xpatchcmd{\proof}{\hskip\labelsep}{\hskip5\labelsep}{}{}
\usepackage{balance}
\usepackage{mathtools}

\usepackage{array}

\begin{document}
\title{Satellite Clustering for Non-Terrestrial Networks: Concept, Architectures, and Applications}

\author{Dong-Hyun Jung, Gyeongrae Im, Joon-Gyu Ryu, Seungkeun Park, Heejung Yu, and Junil Choi\\
\thanks{\textit{Dong-Hyun Jung, Gyeongrae Im, Joon-Gyu Ryu, and Seungkeun Park are with the Radio \& Satellite Division, Electronics and Telecommunications Research Institute, Daejeon, South Korea.}}
\thanks{\textit{Heejung Yu (corresponding author) is with the Department of Electronics and Information Engineering, Korea University, Sejong, South Korea.}}
\thanks{\textit{Junil Choi (corresponding author) is with the School of Electrical Engineering, KAIST, Daejeon, South Korea.}}
\vspace{-0.7cm}}

\maketitle

\begin{abstract}
Recently, mega-constellations with a massive number of low Earth orbit (LEO) satellites are being considered as a possible solution for providing global coverage due to relatively low latency and high throughput compared to geosynchronous orbit satellites. However, as the number of satellites and operators participating in the LEO constellation increases, inter-satellite interference will become more severe, which may yield marginal improvement or even decrement in network throughput. In this article, we introduce the concept of satellite clusters that can enhance network performance through satellites' cooperative transmissions. The characteristics, formation types, and transmission schemes for the satellite clusters are highlighted. Simulation results evaluate the impact of clustering from coverage and capacity perspectives, showing that when the number of satellites is large, the performance of clustered networks outperforms the unclustered ones. The viable network architectures of the satellite cluster are proposed based on the 3GPP standard.  Finally, the future applications of clustered satellite networks are discussed.
\\
\end{abstract}

\vspace{-0.5cm}
\IEEEpeerreviewmaketitle

\section{Introduction}
Satellite communications have been recently employed to provide global internet services exploiting the large coverage of satellites. 
The beam size of geosynchronous orbit (GEO) satellites is typically 200-3500 km, while that of non-GEO satellites such as medium and low Earth orbits is 100-1000 km [\ref{Ref:3GPP_38.821}].
Although the low Earth orbit (LEO) satellites' coverage is small compared to that of GEO satellites, the LEO satellites have recently received great attention because of relatively low latency and high throughput due to their low altitude. In addition, the LEO satellites are becoming more miniaturized in size, integrated, and light-weighted, which reduces manufacturing time and launch costs [\ref{Ref:Radhakrishnan}].

Typically, as the number of satellites in the constellation increases, the throughput of the satellite network is also enhanced. However, when a sufficiently large number of satellites has already been launched, adding more satellites in orbit may increase inter-satellite interference, resulting in the marginal enhancement of network throughput. Motivated by this, the concept of \textit{satellite cluster} (i.e., a set of closely-located small satellites) has been investigated to further enhance the performance through cooperation [\ref{Ref:Liu}].

\begin{figure*}[!t]
\begin{center}
\includegraphics[width=1.8\columnwidth]{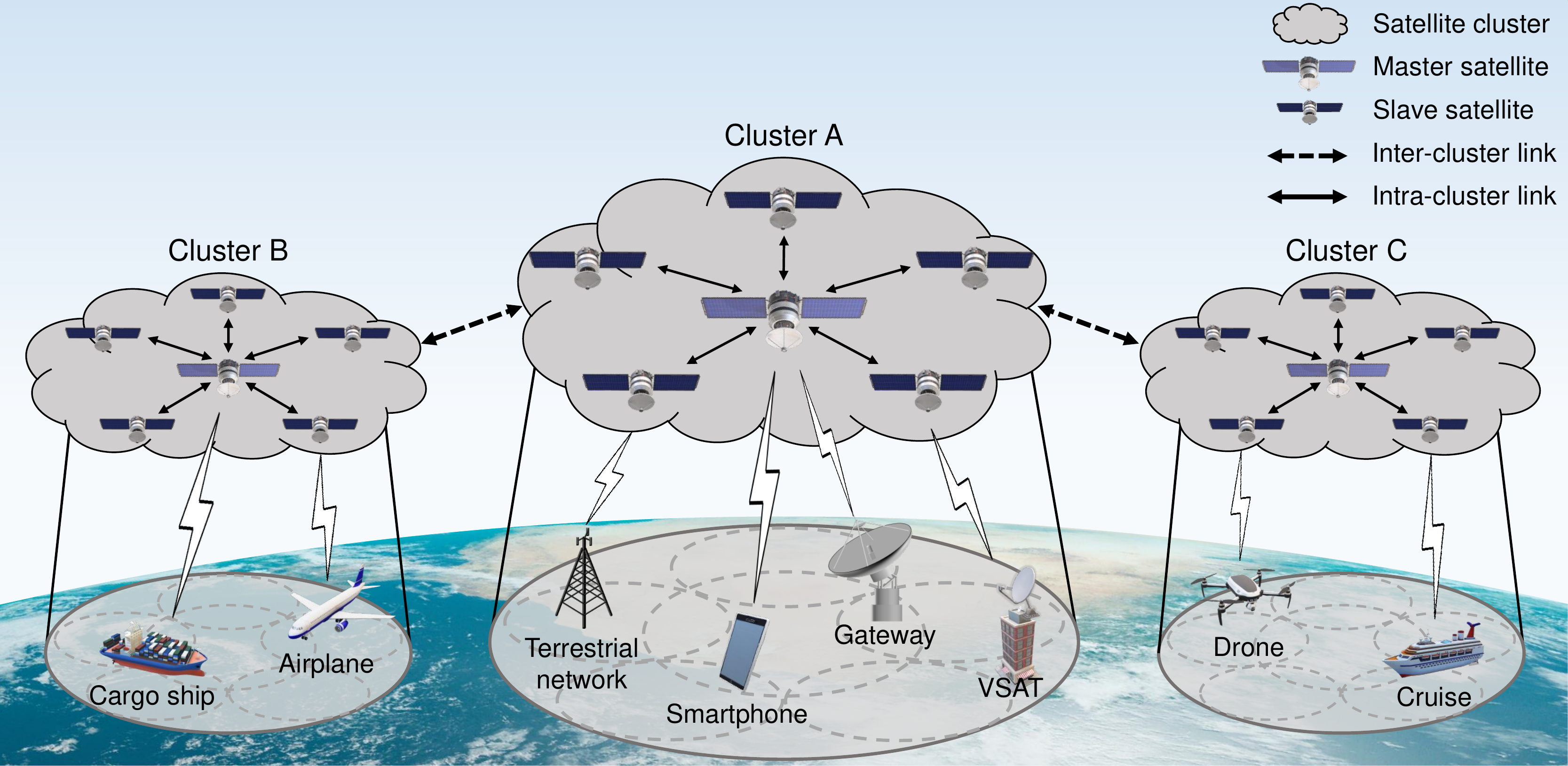}
\end{center}
\setlength\abovecaptionskip{.25ex plus .125ex minus .125ex}
\setlength\belowcaptionskip{.25ex plus .125ex minus .125ex}
\caption{Concept of the satellite clusters.}
\label{Fig:system_model}
\end{figure*}

A satellite cluster is a group of multiple satellites placed nearby where the satellites cooperatively transmit and receive signals as if they were a single multi-antenna satellite as shown in Fig.~\ref{Fig:system_model}. The satellites in a cluster act as either master or slave according to the pre-assigned roles where the cluster consists of one master and multiple slave satellites. The master satellite manages the entire cluster and plays a key role in cooperative transmissions by exchanging information and control signals with the slave satellites through inter-satellite links (ISLs). Thus, powerful on-board processors for inter-satellite communications and routing are required at the master's payload. The slave satellites passively operate as directed by their master and serve as cooperative nodes to improve the performance of the cluster. 

In this article, we investigate satellite clusters to enhance the network capacity by cooperative transmissions among multiple satellites. We describe the characteristics, formation types, and transmission schemes of the satellite clusters and evaluate the network performance from coverage and capacity perspectives by stochastic geometry-based simulations. A key finding is that when the number of satellites is large, the performance of the clustered satellite networks is better than that of the unclustered networks. We also propose and compare 3GPP-based architectures for the satellite clusters and address possible challenges. The applications of the satellite clusters are also discussed.

\section{Clustered Satellite Networks}
In this section, we discuss the satellite cluster's characteristics, formation, and cooperative transmission schemes. We also evaluate the network performance in terms of capacity and coverage probability based on the stochastic geometry.

\subsection{Benefits}
Different from the unclustered satellite constellations, the satellite clusters have the following advantages.

\subsubsection{Geographical Proximity Among Satellites}
The satellites in a cluster shape a certain formation and move in groups.
Thus, the relative distances among the satellites are very close, e.g., several to hundreds of kilometers according to the cluster size.
Thanks to this geographical proximity, accurate beam pointing between satellites, which is generally considered as a main challenge of conventional ISLs, becomes more straightforward because the distance between the master and slave is much shorter than that of the conventional ISLs, e.g., hundreds of kilometers for LEO-to-LEO ISLs while tens of thousands of kilometers for GEO-to-GEO ISLs. Moreover, as the orbits of slaves are determined by the master's orbit, adding more slaves in a cluster requires negligible orbital resources.

\subsubsection{Simple Payload and Scalability}
To cope with increased user requirements, satellites should have more complex functionalities. This in turn causes larger payloads and higher production and launch costs.
The satellite cluster can solve this problem by dividing functionalities of one complex master satellite into multiple smaller slave satellites.
Thus, the slave satellites may have as few functionalities as possible and take a portion of the master's load.
Moreover, as the slave satellites cannot operate without the master but are used only for coverage extension and throughput increase, the full protocol stack is not a requirement for the slaves.
These make it easy to reduce the payload of the master as well as scale up the cluster.

\subsubsection{Spatial Diversity}
The LEO satellites do not always have a direct path to a ground terminal due to their low altitude since the probability that the satellites experience a line-of-sight (LOS) channel varies according to the elevation angle [\ref{Ref:3GPP_38.811}].
Transmissions from multiple satellites in the cluster may increase the LOS probability and overcome the performance degradation by blockages. 
Although a dominant LOS path between one satellite and a ground terminal exists, the channels from the satellites in a cluster to the terminal may not be highly-correlated because the distance between satellites is much farther than the wavelength of the carrier frequency, e.g., several to hundreds of kilometers. Therefore, spatial diversity can be achieved, and this would enhance the network throughput [\ref{Ref:Yu2}].

\subsubsection{Coverage Maintainability}
When the satellites are in operation after settling in orbit, some satellites may break down due to unexpected hardware or software problems. In the unclustered constellations, if a satellite malfunctions or does not work for a certain time, its coverage area can be served by spare satellites in the same or another orbital plane. However, if the ISL capability of the malfunctioning satellite is also out of order, the inter-satellite handover procedures cannot be appropriately initiated. Moreover, when the malfunctioning satellite is replaced by one in another orbital plane, these satellites should support the inter-plane ISLs for handovers, which results in huge burdens for the beam alignment and on-board routing. On the contrary, in the clustered constellations, the coverage is hardly reduced even if one of the slave satellites malfunctions because the other satellites in the cluster still can provide services without handovers.

\subsection{Cluster Formation}
\begin{figure}
\centering
\subfigure[]{
\includegraphics[width=\columnwidth]{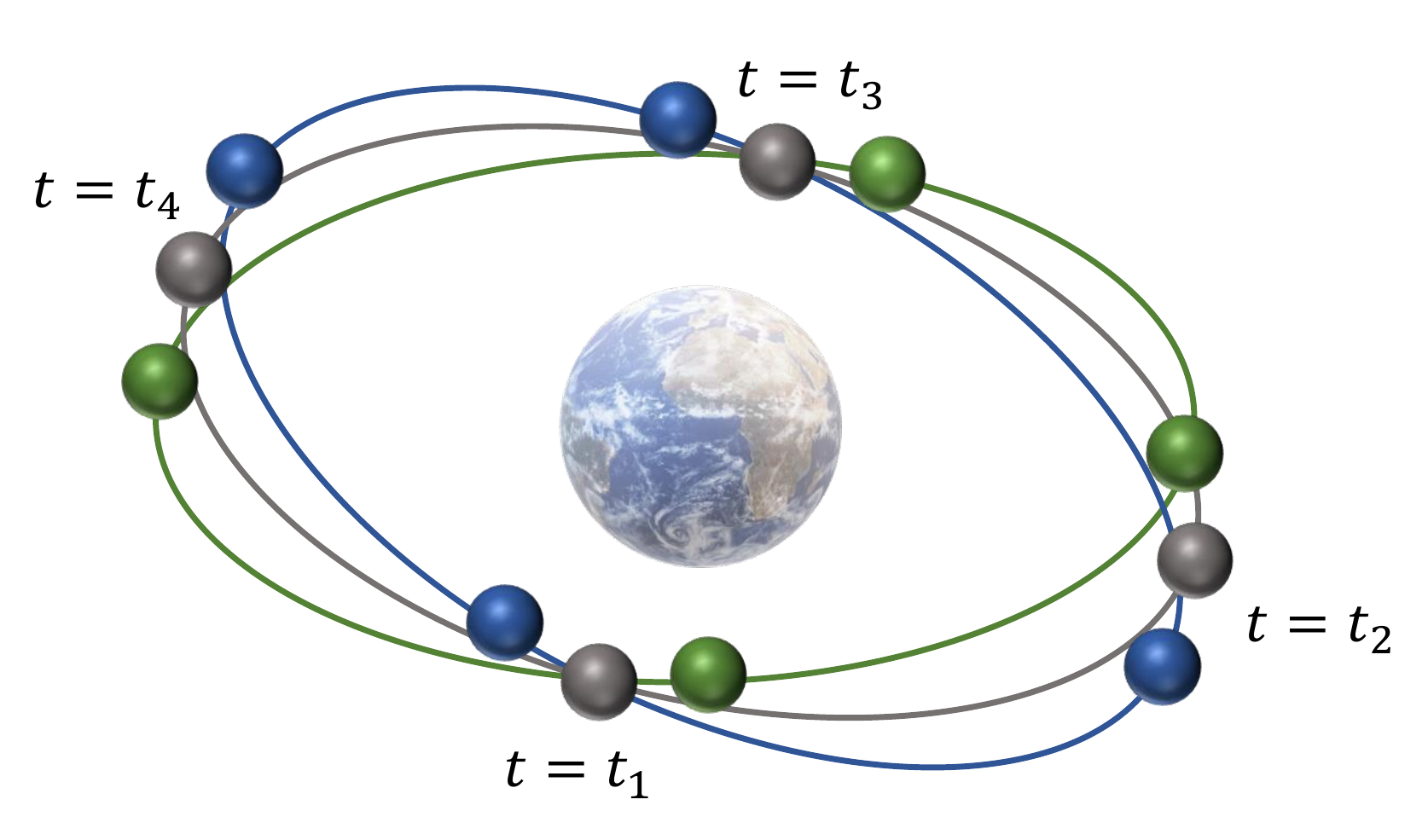}
\label{Fig:PCO_a}
}
\hspace{5cm}
\subfigure[]{
\includegraphics[width=\columnwidth]{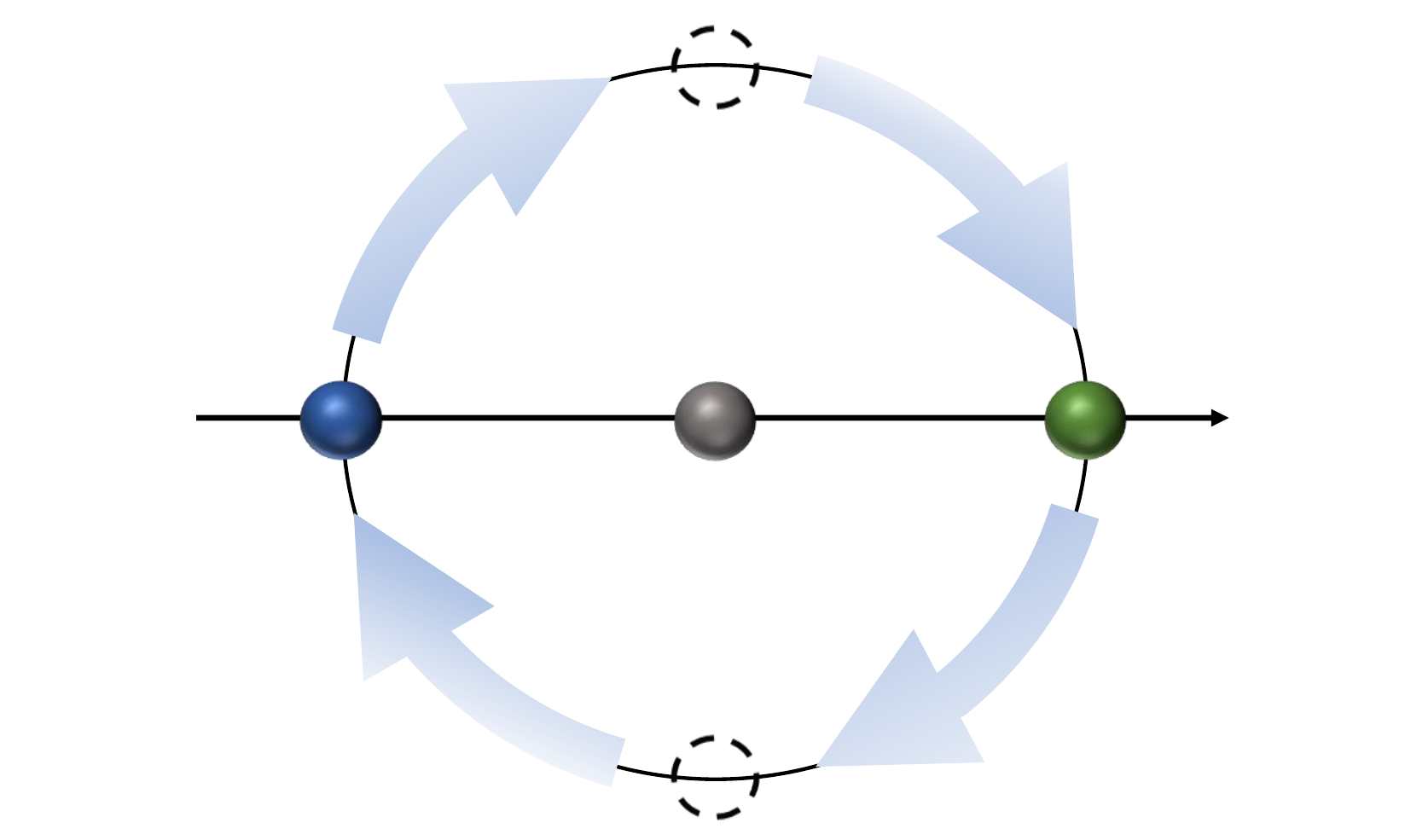}
\label{Fig:PCO_b}
}
\caption{(a) Orbital configuration and (b) view from the Earth for the PCO-based satellite cluster with three satellites where the gray sphere is the master satellite, and the blue and green are the slaves. From $t=t_i$ to $t=t_{i+1}$, $i=\{1,2,3\}$, the slaves rotate 90 degrees when viewed from the Earth [\ref{Ref:Eyer}].}
\label{Fig:PCO}
\end{figure}

A possible candidate to enable formation flying of satellites is the \textit{projected circular orbit (PCO)}. In the PCO-based formation flying, the master satellite in the cluster follows a reference orbit, while the remaining slave satellites travel along the PCOs, which have slightly different inclinations and eccentricities from the reference orbit  as shown in Fig. \ref{Fig:PCO}. With this small change of the orbital configuration, the slave satellites tend to circularly orbit the master satellite when viewed from the Earth.
This PCO-based formation flying was successfully demonstrated in the Canadian nanosatellite program called CanX-4\&5 mission [\ref{Ref:Eyer}]. By simply extending this concept to multiple satellites, the master satellite surrounded by more than two slave satellites could be implemented [\ref{Ref:Popov}]. In this article, we consider two PCO-based cluster formations: circular and uniform clusters, as shown in Fig. \ref{Fig:cluster_dist}.

\subsubsection{Circular Cluster}
A circular cluster is a cluster in which the slave satellites are equidistant from the master satellite, forming a circular swarm. With the circular formation, the distances of the ISLs are comparable, which is a big advantage for synchronizing transmission timing across the satellites in the cluster. In addition, when the slave satellites are equally distant from the adjacent slaves, it is expected that the antenna or lens alignment for ISL communications can be simple.
However, when the number of satellites in clusters is large, there exists a space limit to deploy the slave satellites over a ring-shaped track.

\subsubsection{Uniform Cluster}
In a uniform cluster, the master satellite is in the middle of the spherical cap where the slave satellites are uniformly distributed. This formation may be made up of multi-layered PCOs with different distance to the master satellite [\ref{Ref:Popov}].
In contrast to the circular formation, the uniform cluster has unequal distances of the ISLs, which yields difficulty in timing synchronization of the satellites and transceiver alignment of the ISLs. This formation, however, is appropriate for dense deployment of satellites, since the uniform distribution efficiently uses the distributed area. In other words, the uniform formation can accommodate more satellites than the circular formation, when the two formations have the same minimum distance among the satellites.

\begin{figure}
\begin{center}
\includegraphics[width=\columnwidth]{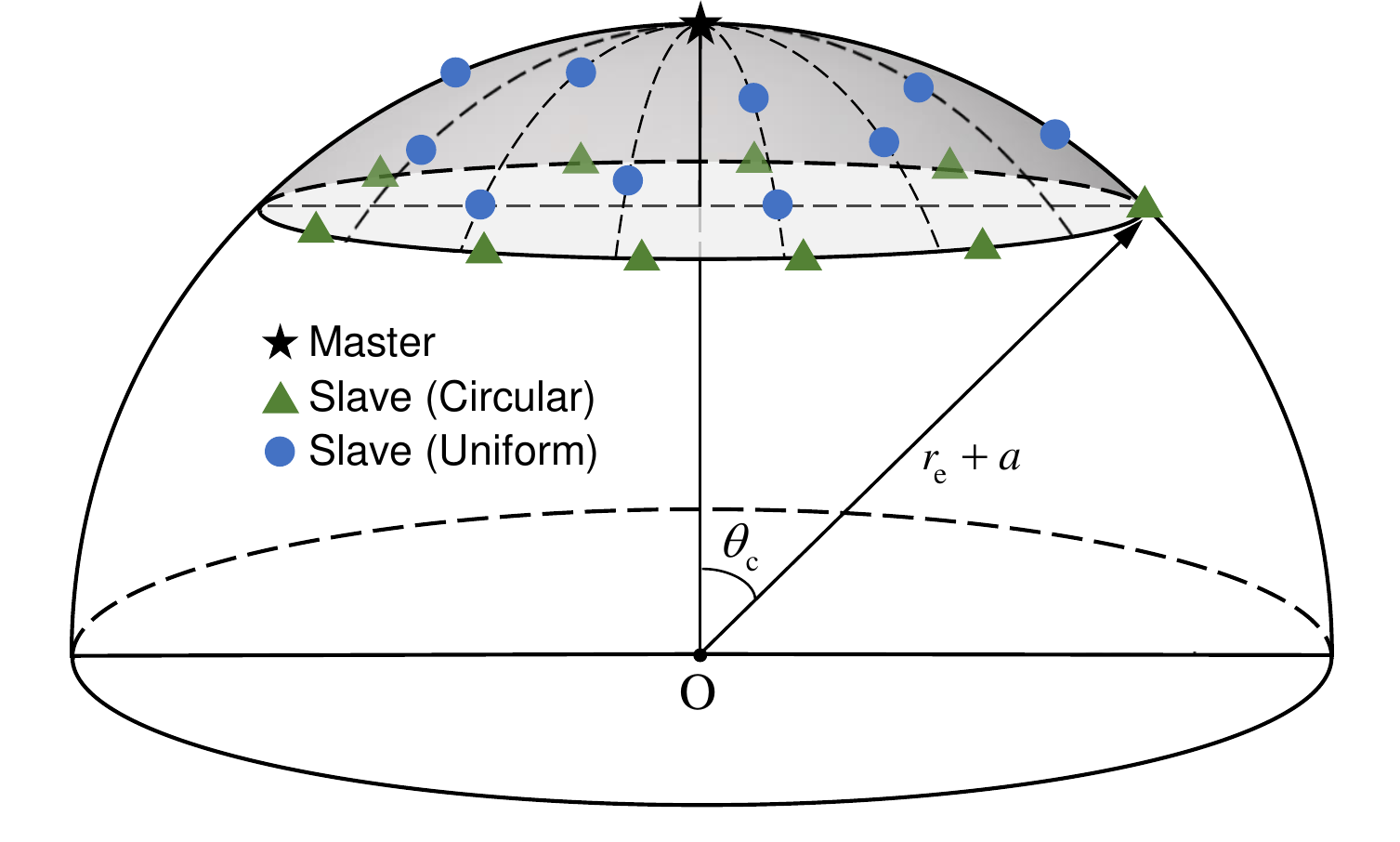}
\end{center}
\setlength\abovecaptionskip{.25ex plus .125ex minus .125ex}
\setlength\belowcaptionskip{.25ex plus .125ex minus .125ex}
\caption{Circular and uniform cluster distributions. The shaded area is the spherical-cap-shaped region where the satellites in a cluster can be distributed. The altitude, the radius of the Earth, and the polar angle of the spherical cap are denoted by $a$, $r_\mathrm{e}$, and $\theta_\mathrm{c}$, respectively.}
\label{Fig:cluster_dist}
\end{figure}

\subsection{Cooperative transmissions}
The concept of cooperative multi-point (CoMP) was proposed to enhance the throughput of cell-edge users by mitigating inter-cell interference from multiple transmission points. 
Several scenarios for intra-eNB CoMP, which uses multiple remote radio heads (RRHs) to perform CoMP at a single eNB, were considered in Release 11 assuming ideal backhauling, while Release 12 focused on inter-eNB CoMP, i.e., CoMP involving multiple eNBs, with non-ideal backhauling. The 3GPP has considered several transmission schemes for downlink CoMP such as joint transmission and dynamic point selection, which can be applied to the satellite clusters.

\subsubsection{Joint Transmission (JT)}
The JT uses multiple transmission points in a cell (intra-cell JT) or in different cells (inter-cell JT) to transmit signals to a user. 
The maximum ratio transmission (MRT) can be applicable to the JT of the satellite clusters.
However, as the amplitude and phase of the channel coefficients are required for the MRT, which should be first estimated and then distributed to all the slave satellites, strict requirements of the ISL would be necessary. 
In addition, the satellites may have a problem in amplifying signals because the satellites usually operate near the saturation level of power amplifier to compensate for the large path-loss.
This results in performance degradation due to non-linear distortion of the desired signals. Instead of the MRT, the equal gain transmission can be used as an alternative for the JT, equally allocating transmit power to the satellites in the cluster. This scheme only requires the phase of channel coefficients, which allows the inexpensive power amplifiers to be mounted at the satellite's payload. 
This can be a great merit for the satellite clusters because the power amplifier of the satellite payload (e.g., traveling wave tube amplifier) is a key component to compensate for the substantial power loss of the received signals.
However, as the JT relies on precise cooperation among satellites in the cluster, strict requirements on the ISL capacity and synchronization are necessary especially  with a large number of slaves.

\subsubsection{Dynamic Point Selection (DPS)}
The DPS is a very simple beamforming scheme that selects only a single transmission point with the best channel condition.
The DPS can mitigate the inter-cell interference because it mutes other transmission points that are not selected.
With this advantage, the DPS would be well applicable to the clustered satellite networks, especially with a massive number of satellites.
Unless the master is the best choice, the master needs to inform only one slave, which has the best channel condition, through the intra-cluster link. In this regard, the DPS can significantly reduce the requirement for ISL capacity and timing synchronization among the satellites.
However, the DPS leads to frequent changes between satellites in the cluster, so low-latency ISL switching is required.

\subsection{Performance Evaluation}
Considering the discussion so far, we evaluate the downlink performance of clustered satellite networks using stochastic geometry-based simulations [\ref{Ref:Wang}]. Assume that the satellites are located at the altitude of 600 km and operated in S-band (2 GHz) with 30 MHz bandwidth. The free-space path-loss model is adapted with the path-loss exponent of~3, and the shadowed-Rician fading is assumed with average shadowing.
For simplicity, we assume the satellites generate a single beam with the maximum transmit antenna gain of 30 dBi, the 3-dB beamwidth of 20 degrees, and the beam pattern given in [\ref{Ref:Chu}]. The satellites maintain their beam boresight in the direction of the subsatellite point.
The terminal has an omni-directional antenna with the gain of 0 dB.
The EIRP density and noise spectral density are set to 34 dBW/Hz and $-174$ dBm/Hz, respectively.
For the clustered networks, the polar angle of the spherical cap where the satellites can be distributed is set to 1 degree and 10 percent of the total satellites are the masters, while the remaining satellites are the slaves. For example, when the number of satellites is 1000, there are 100 clusters, each consisting of one master and nine slaves.
For the unclustered networks, all satellites are independently distributed where each satellite works as an independent transmitter.

We use a homogeneous binomial point process (BPP) to model the cluster distribution [\ref{Ref:Wang}].
The masters are uniformly distributed according to the homogeneous BPP over a sphere.
For the circular formation, the slaves are spaced at the boundary of the spherical cap, as shown in Fig. \ref{Fig:cluster_dist}, maintaining relatively equal distances to the adjacent slaves. For the uniform formation, the slaves are distributed on the spherical cap according to the homogeneous BPP.

\begin{figure}
\begin{center}
\includegraphics[width=0.85\columnwidth]{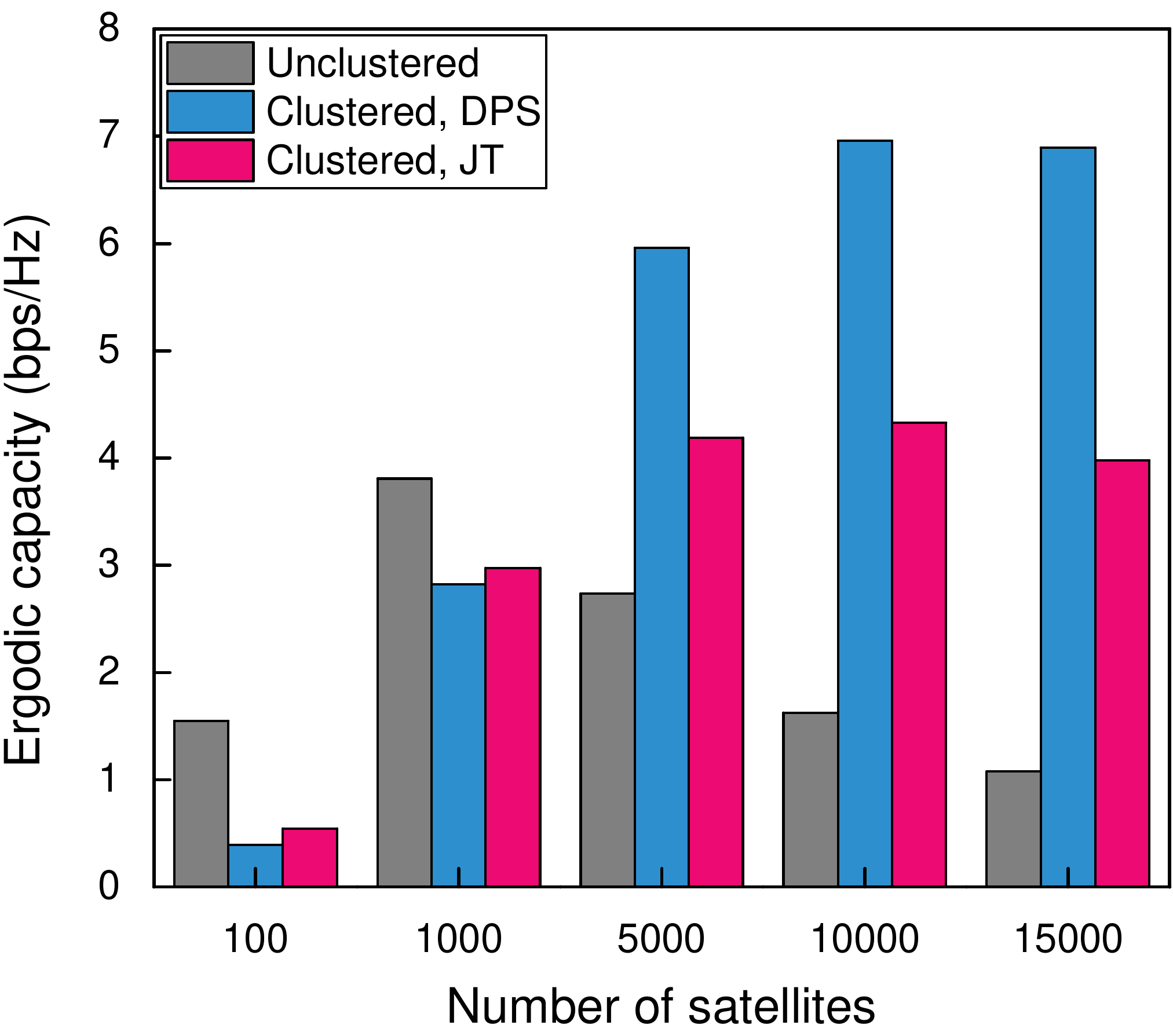}
\end{center}
\setlength\abovecaptionskip{.25ex plus .125ex minus .125ex}
\setlength\belowcaptionskip{.25ex plus .125ex minus .125ex}
\caption{Ergodic capacity versus the number of satellites.}
\label{Fig:Cerg_vs_N}
\end{figure}

In Fig. \ref{Fig:Cerg_vs_N}, the ergodic capacities of the unclustered and clustered networks are compared with various numbers of satellites, assuming the circular formation and the MRT-based JT for the clustered networks.
When the number of satellites is small, the unclustered network achieves a higher ergodic capacity than the clustered networks.
In contrast, for large numbers of satellites, the satellite cluster achieves higher performance by cooperative transmissions.
This proves that the satellite cluster can play an important role to enable mega-constellations as more satellites exist in the space.
With an excessively large number of satellites, however, the performance of both unclustered and clustered networks decreases due to the higher inter-satellite interference.
This explains that the constellation design with an appropriate number of satellites is very important in terms of the network performance. For different beamwidths, this tendency still holds, but the number of satellites at which the performance starts to degrade would be increased with a smaller beamwidth and vice versa.
The JT has a greater capacity than the DPS for small numbers of satellites due to the optimality of the MRT with negligible inter-satellite interference.
In contrast, the DPS outperforms the JT when the number of satellites becomes large because the DPS significantly reduce the inter-satellite interference.

\begin{figure}
\begin{center}
\includegraphics[width=0.85\columnwidth]{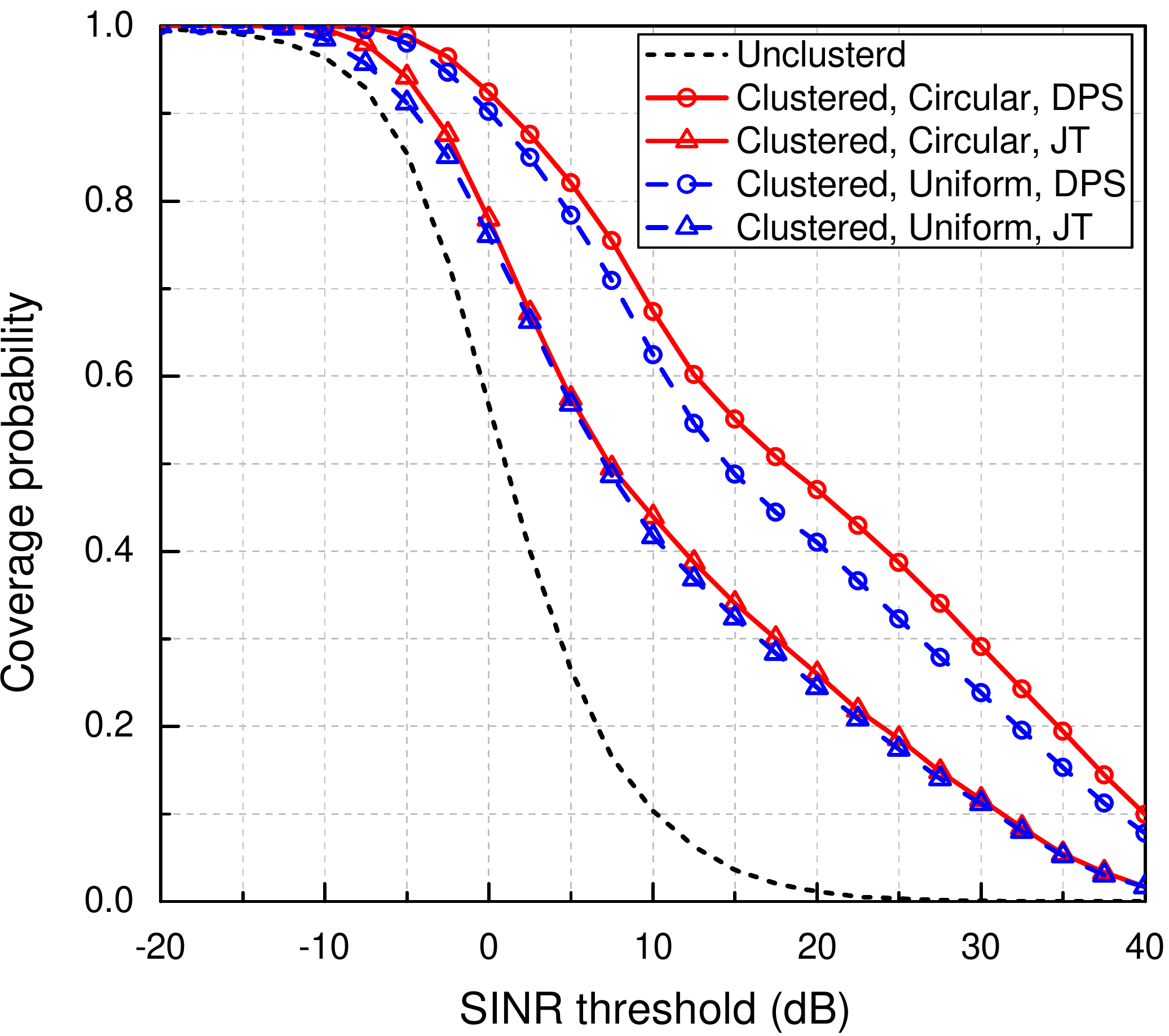}
\end{center}
\setlength\abovecaptionskip{.25ex plus .125ex minus .125ex}
\setlength\belowcaptionskip{.25ex plus .125ex minus .125ex}
\caption{Coverage probability versus SINR threshold when the total number of satellites is 10k.}
\label{Fig:Pcov_vs_beta}
\end{figure}

Fig. \ref{Fig:Pcov_vs_beta} shows the coverage probabilities, i.e., the probabilities that the signal-to-interference-plus-noise ratio (SINR) is higher than a threshold.
As expected, the clustering has benefits in terms of coverage probability. 
It is shown that the satellite clusters with the circular formation have better performance than that with the uniform formation.
Thus, with the small cluster size, the circular formation is preferable due to low requirements for ISLs and accuracy of position control, while uniform formation may be suitable with the large cluster size because the satellites can be deployed efficiently.

\begin{center}
\begin{table*}[ht]
\caption{Candidate Options for Functional Split}\label{Table:Options}
{\small
\hfill{}
\resizebox{\textwidth}{!}
{\begin{tabular}{|c|l|l|l|l|}
\hline
\multirow{2}{*}{Options}
& \multicolumn{1}{c|}{\multirow{2}{*}
{Requirements}}
& \multicolumn{1}{c|}{\multirow{2}{*}
{Pros}}
& \multicolumn{1}{c|}{\multirow{2}{*}
{Cons}}
& \multicolumn{1}{c|}{\multirow{2}{*}{
\begin{tabular}[c]{@{}c@{}}
Applicable satellite 
environments
\end{tabular}}}\\
& \multicolumn{1}{c|}{}
& \multicolumn{1}{c|}{}
& \multicolumn{1}{c|}{}
& \multicolumn{1}{c|}{}\\ \hline \hline
\begin{tabular}[c]{@{}c@{}}
Intra-PHY split
\end{tabular} 
& \begin{tabular}[c]{@{}l@{}}
\tabitem UL data rate: 86.1 Gbps\\
\tabitem  DL data rate: 86.1 Gbps\\
\tabitem Latency: $\sim$100 us\\
\end{tabular}   
& \begin{tabular}[c]{@{}l@{}}
\tabitem Low installation cost\\
\tabitem Cost-effective RRH\\
\tabitem Ideal for CoMP
\end{tabular}                           
& \begin{tabular}[c]{@{}l@{}}
\tabitem  High fronthaul capacity requirement\\
\tabitem Strict latency requirement\\
\tabitem Subframe-level timing \\\hspace{3.4mm} between CU and DU
\end{tabular} 
& \begin{tabular}[c]{@{}l@{}}
\tabitem  Centralized architecture\\
\tabitem High ISL capacity\\
\tabitem High data rate requirements\\
\tabitem Virtual-RAN
\end{tabular}  \\ \hline
\begin{tabular}[c]{@{}c@{}}
Intra-MAC split
\end{tabular} 
& \begin{tabular}[c]{@{}l@{}}
\tabitem  UL data rate: 3 Gbps\\
\tabitem  DL data rate: 4 Gbps\\
\tabitem Latency: $\sim$1 ms\\
\end{tabular}    
& \begin{tabular}[c]{@{}l@{}}
\tabitem  Low fronthaul capacity requirement\\
\tabitem Low HARQ buffer requirement \\\hspace{3.4mm} in master\\
\tabitem Low HARQ latency\\
\tabitem CoMP possible\end{tabular} 
& \begin{tabular}[c]{@{}l@{}}
\tabitem  Scheduling complexity \\\hspace{3.4mm} between CU and DU\\
\tabitem Limitations for \\\hspace{3.4mm} some CoMP schemes
\end{tabular}                          
& \begin{tabular}[c]{@{}l@{}}
\tabitem  Low bandwidth requirement\\
\tabitem Enhancing reliability via HARQ
\end{tabular}\\ \hline
\begin{tabular}[c]{@{}c@{}}
PDCP/RLC split 
\end{tabular}  
& \begin{tabular}[c]{@{}l@{}} 
\tabitem UL data rate: 3 Gbps\\ 
\tabitem DL data rate: 4 Gbps\\ 
\tabitem Latency: 1$\sim$10 ms\\ 
\end{tabular} 
& \begin{tabular}[c]{@{}l@{}}
\tabitem  Traffic offloading\\
\tabitem Existing DC available\\
\tabitem Low fronthaul capacity requirement\\
\tabitem Low latency requirement
\end{tabular}  
& \begin{tabular}[c]{@{}l@{}}
\tabitem  Security issue in PDCP\\
\tabitem Limitations in \\\hspace{3.4mm} cooperative functions
\end{tabular}                        
& \begin{tabular}[c]{@{}l@{}}
\tabitem  Latency tolerant\\
\tabitem Long distance between master/slave\\
\tabitem Imperfect/limited ISL
\end{tabular} \\ \hline
\end{tabular}}
\hfill{}
}
\end{table*}
\end{center}

\section{Practical Architectures}
In this section, we propose the 3GPP-based network architectures for the satellite clusters considering functional split options presented in [\ref{Ref:3GPP_38.801}]. We assume that the master has all the network functions of gNB in 5G new~radio (NR) and is connected to the 5G core (5GC) network through a gateway using the NG interface logically and the satellite radio interface physically.
We discuss feasible functional split options that can be applicable to the clustered satellite networks and address several technical challenges to enable the proposed architectures for the satellite clusters.

\subsection{Functional Splits in 3GPP}
A centralized or cloud radio access network (C-RAN) architecture has been proposed to increase the installation efficiency of base stations in a cost-effective manner. The C-RAN physically separates the base station into the remote radio head (RRH) and baseband unit (BBU). The RRHs are deployed and distributed at each cell site while the BBUs are co-located and centralized. The Common Public Radio Interface is used as the fronthaul interface between the RRH and BBU. With such a C-RAN architecture, the centralized BBU is able to reduce the rental cost and electricity rate by performing baseband processing for the multiple distributed RRHs.

In 5G NR, however, the C-RAN structure brings another problem that the fronthaul requires a much higher capacity to transmit the I/Q data sampled in the time domain. In order to mitigate such a problem, the open RAN structure has been proposed by selectively applying eight functional split options for the gNB [\ref{Ref:3GPP_38.801}]. Note that, in 3GPP, Option 2, i.e., the PDCP/RLC split, was recommended as the higher layer split, while the open RAN standard selected Option 7-2x, i.e., the low-PHY split between the resource element mapper and beamformer, as the lower layer split. Different from the C-RAN, the open RAN splits the functions of the gNB in higher layers into two units: distributed unit (DU) and centralized unit (CU). 

\subsection{Split Options for Satellite Cluster}
Specifically, the master is assumed to have full functionalities of the gNB (DU and CU), while the only gNB-DU is equipped at each slave to support the cluster's cooperative transmissions.
With this assumption, we propose the following three functional split options between the master and slave: 
\begin{itemize}
    \item Intra-physical layer (intra-PHY) split
    \item Intra-medium access control layer (intra-MAC) split
    \item Packet data convergence protocol/radio link control (PDCP/RLC) split.
\end{itemize}
The features and applicability of the three options are summarized in Table \ref{Table:Options} where the required data rates are calculated with the following parameters: 100 MHz bandwidth, 256-quadrature amplitude modulation, 32 antenna ports, and 8 multiple-input multiple-output layers~[\ref{Ref:ITU}]. 
Satellite operators planning to offer 3GPP-based services with satellite clusters may select the best split option based on the following discussions.

\subsubsection{Intra-PHY Split} 
\begin{figure*}[!t]
\begin{center}
\includegraphics[width=1.9\columnwidth]{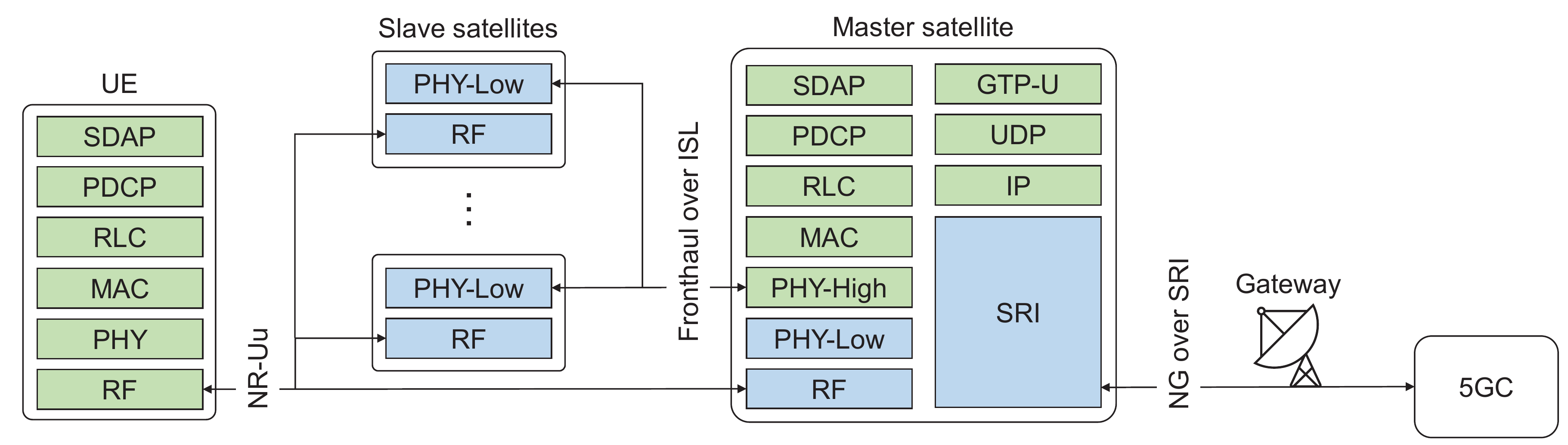}
\end{center}
\setlength\abovecaptionskip{.25ex plus .125ex minus .125ex}
\setlength\belowcaptionskip{.25ex plus .125ex minus .125ex}
\caption{Cluster architecture with intra-PHY split. (SDAP: service data adaptation protocol, IP: Internet protocol, UDP: user datagram protocol, GTP-U: general packet radio service tunnelling protocol-user plane)}
\label{Fig:ORAN}
\end{figure*}

As shown in Fig. \ref{Fig:ORAN}, in this split option, the slave has the low-PHY layer, including orthogonal frequency division multiplexing modulation/demodulation and resource element mapping/demapping, and the RF part, while the master includes the other upper-layer functions.
This intra-PHY split is similar to the open RAN structure, and it can be a suitable option to ensure compatibility and coordination with the terrestrial network (TN). This option suggests a centralized architecture optimized for utilizing NR features, such as CoMP and carrier aggregation (CA). 
It also requires the simplest architecture among the three considered options for the slave satellite's payload. This advantage significantly reduces the manufacturing and launch costs of the slaves and saves the energy for signal processing, which consequently makes it easy to scale up the number of satellites to obtain more diversity gain.
However, the data rate requirement in the ISL significantly increases and the latency requirement must be very low.
Thus, this option can be applied only when the distance between the master and slave is close enough to satisfy the latency requirements, and the ISL between master and slave satellites is nearly perfect.

\subsubsection{Intra-MAC Split} This option splits the functions in the middle of MAC layer, that is, radio resource control (RRC), PDCP, RLC, and high-MAC layers are in the master, and low-MAC and PHY layers as well as the RF part are in the slave. This intra-MAC split is aimed at reducing the latency of hybrid automatic repeat request (HARQ) protocol to ease the constraints of fronthaul capacity and increase reliability while ensuring data rate. Since HARQ is processed in DU, the slaves handle their own HARQ processes.
Hence, the master can have much smaller buffer and less complex processor because the master only handles its own HARQ. Still, some of CoMP functions and CA can be utilized at the master. 
However, the interface between CU/DU becomes complex, and the scheduling operations over CU/DU should be defined additionally.

\subsubsection{PDCP/RLC Split} In this option, RRC and PDCP layers are in the master, and RLC, MAC, and PHY layers and the RF part are in the slave.
With this higher-layer split, the slaves take a large portion of gNB's processing, resulting in the low ISL capacity and latency requirements.
The clustered satellite network with this split can utilize dual connectivity (DC) since it has a similar structure to the existing DC.
When the cluster size is large, the satellites may have a large delay and performance degradation in the ISLs due to the long distance and antenna or lens misalignment. 
Thus, this option is well applicable to the satellite cluster especially when the satellites are far apart.
On the contrary, the network functions up to the RLC layer should be implemented at the slaves, which requires complex payloads and reduces the scalability of the cluster.
In addition, it is difficult to use cooperative transmission schemes such as CoMP since MAC layer is not centralized, and there may be security issues because the coordination of security configurations between different PDCP instances is required.

\subsection{Related Challenges}
\subsubsection{Satellite Position Control}
Maintaining the formation of satellites in a cluster is crucial to guarantee reliable inter-satellite communications. 
However, the accurate position control of the multiple satellites is challenging due to environmental disturbances such as atmospheric drag and solar pressure [\ref{Ref:Bandyopadhyay}].
This may require a complex ground control system as well as high-performance sensors on board.
For example, optical sensor-aided position control systems have been developed to enable proximate formation flying of spacecrafts at Marshall Space Flight Center in NASA where the sensors are used to calculate the relative distance between two spacecrafts. Moreover, Goddard enhanced onboard navigation system, also known as GEONS, is a software developed at Goddard Space Flight Center in NASA, which uses standard global positioning system (GPS) receivers and onboard sensors to provide accurate relative navigation solutions in real time.

\subsubsection{Limited Bandwidth of ISL}
In order to enable the proposed architectures for the satellite cluster, the high data rate is required at the ISL as discussed in Table \ref{Table:Options}.
Especially, the intra-PHY split requires the data rate in the ISL up to approximately 86 Gbps.
As a solution to achieve such high data rate, free-space optical (FSO) communications can be used for the ISLs instead of the RF interface.
However, several challenges must be addressed to apply the FSO for ISLs between the master and slave satellites [\ref{Ref:Halim}]. 
As slightly different orbits between the master and slave make the difference in relative angular motion of two satellites, a point-ahead angle prediction is required to compensate for the difference. In addition, due to the narrow beamwidth of the FSO and large ISL distance, the algorithms for acquisition, tracking, and pointing must be accurate. 
Since the capacity of the current FSO technology may not satisfy the fronthaul data rate requirement of the Intra-PHY split, the compression of fronthaul data can be a promising approach for the proposed architecture [\ref{Ref:Yu}].

\subsubsection{Synchronization}
Time and frequency synchronization is a critical issue for the satellite cluster.
The signals from the satellites may experience different propagation delays caused by the following factors: 
\begin{itemize}
    \item Different distances between the UE and the satellites in the cluster
    \item Different distances between the master and slaves (for uniform clusters).
\end{itemize}
In addition, the relative velocity of the satellites in the cluster may be dissimilar due to unequal elevation angles, which causes different Doppler shifts in the signals from the satellites.
This misalignment in time and frequency would degrade the performance of the cooperative transmissions.
The pre-compensation for the timing difference and Doppler shift can be done to resolve this time and frequency uncertainty. For example, the timing and frequency offsets can be pre-compensated by obtaining the position of the UE through GNSS and the speed and position of the satellite through the ephemeris information [\ref{Ref:3GPP_38.821}].

\subsubsection{Master Dependency} 
As the master plays an important role in the operation of the cluster, the slaves are highly dependent on the master. In the worst case the master breaks down, the loss of the master would render the whole cluster unusable, which is the inherent problem due to the master-slave relation. To mitigate this problem, multiple masters may be placed in the cluster. The additional master may work as a slave in the normal operation or remain inactive as a spare master. If the original master fails, the additional master would start to serve as a real master.

\section{Future Applications}
In this section, we discuss possible application scenarios where the merits of the satellite clusters for LEO mega-constellations can be exploited.
\subsubsection{Direct Access to Smartphones}
To expand application scenarios of satellite communications, direct access to smartphones is an essential requirement. Compared to the conventional UE in satellite communications, smartphones may have a lower antenna gain, for example, up to 5 dB less gain, and lower transmit power due to less battery capacity. Therefore, the enhancement of the link budget is required for the direct access. Recent trials for direct access to smartphones include iPhone14's satellite communication feature and the cooperation between SpaceX and T-Mobile, but both provide low data rate services in some limited applications. The cooperative transmissions among clustered satellites can be one of the promising approaches to this end.

\subsubsection{Distributed Computing}
As various services demanding extremely high computational loads emerge, distributed computing is a key requirement in future networks.
The satellites in a cluster may work as edge nodes performing distributed computing. For example, the slaves may offload the computation tasks of the master in a distributed manner.
With the pre-configured topology of the cluster, stable and expectable computing performance can be achieved. With unclustered satellites, however, dynamic group formation and ISL management are required to incorporate multiple LEOs moving independently.  

\subsubsection{Localization}
The satellite clusters can be utilized for localization of UEs instead of the existing GPS. It means that a UE can estimate its position without an additional GPS receiver. Because the LEO satellites are located at much lower altitudes than the GPS satellites, they have the potential to deliver several benefits in terms of navigation, precise point positioning/timing,  and location-enabled communications. With the clustered LEO satellites whose relative positions are maintained, the aforementioned benefits can be more easily obtained than unclustered LEO satellites. This is because the synchronized transmissions of the reference signals among multiple satellites can be simply implemented.
However, the clock accuracy of LEO satellites is very critical for positioning through LEO constellations because errors in clock estimates degrade the accuracy of the GNSS measurements. The LEO satellites can be equipped with low-cost chip-scale atomic clocks to enhance the accuracy.

\subsubsection{Coordination with GEO Networks}
When GEO satellite networks coordinate with the LEO satellite clusters, the GEO satellites can serve the UEs with low requirements on the latency and throughput, while shorter latency and higher throughput are provided by the LEO clusters. This could balance the load between the two satellite networks, and make a better usage of the resources of the whole network.
The DC operation as in 5G NR [\ref{Ref:Monserrat}] could be considered for the LEO and GEO satellite networks to improve throughput and reduce service interruption. For example, the GEOs are configured as the masters, while the LEOs as the slaves.

\section{Conclusions}
In this article, we introduced the concept of a satellite cluster for satellite communications achieving high throughput by cooperative transmissions. The characteristics, formation, and cooperative transmission schemes of the satellite cluster were discussed and stochastic geometry-based simulations were performed to evaluate the coverage and capacity performance. We also proposed the practical architectures for the clustered satellite networks based on the 3GPP standard and discussed the related challenges. The future applications of the clustered LEO satellite networks were also discussed.

\section*{Acknowledgment}
This work was supported by Institute of Information \& communications Technology Planning \& Evaluation (IITP) grant funded by the Korea government (MSIT) (No.2021-0-00847, Development of 3D Spatial Satellite Communications Technology).

\ifCLASSOPTIONcaptionsoff
  \newpage
\fi

\balance

\begin{thebibliography}{1}

\bibitem{bib:3GPP_38.821}\label{Ref:3GPP_38.821}
3GPP TR 38.821 v16.0.0, ``Solutions for NR to support non-terrestrial networks (NTN)," Dec. 2019.

\bibitem{bib:Radhakrishnan}\label{Ref:Radhakrishnan}
R. Radhakrishnan \textit{et al.}, ``Survey of inter-satellite communication for small satellite systems: Physical layer to network layer view," \emph{IEEE Commun. Surveys Tuts.}, vol. 18, no. 4, pp. 2442-2473, 4th Quart. 2016.

\bibitem{bib:Liu}\label{Ref:Liu}
G.-P. Liu and S. Zhang, ``A survey on formation control of small satellites," \emph{Proc. IEEE}, vol. 106, no. 3, pp. 440-457, Mar. 2018.

\bibitem{bib:3GPP_38.811}\label{Ref:3GPP_38.811}
3GPP TR 38.811 v15.4.0, ``Study on NR to support non-terrestrial networks," Sept. 2020.

\bibitem{bib:Yu2}\label{Ref:Yu2}
Q.-Y. Yu \textit{et al}., ``Virtual multi-beamforming for distributed satellite clusters in space information networks," \emph{IEEE Wireless Commun.}, vol. 23, no. 1, pp. 95-101, Feb. 2016.

\bibitem{bib:Eyer}\label{Ref:Eyer}
J. K. Eyer \textit{et al}., ``A formation flying control algorithm for the CanX-4\&5 low Earth orbit nanosatellite mission," \emph{Space Technol.}, vol. 27, no. 4, pp. 147–158, 2007.

\bibitem{bib:Popov}\label{Ref:Popov}
A. M. Popov \textit{et al}., ``Development and simulation of motion control system for small satellites formation," \emph{Electronics}, vol. 10, no. 24, 2021, Art. no. 3111.

\bibitem{bib:Wang}\label{Ref:Wang}
R. Wang, M. A. Kishk and M.-S. Alouini, ``Ultra-dense LEO satellite-based communication systems: A novel modeling technique," \emph{IEEE Communications Magazine}, vol. 60, no. 4, pp. 25-31, Apr. 2022.

\bibitem{bib:Chu}\label{Ref:Chu}
J. Chu \textit{et al}., ``Robust design for NOMA-based multibeam LEO satellite internet of things," \emph{IEEE Internet Things J.}, vol. 8, no. 3, pp. 1959-1970, Feb. 2021.

\bibitem{bib:3GPP_38.801}\label{Ref:3GPP_38.801}
3GPP TR 38.801 v14.0.0, ``Study on new radio access technology: Radio access architecture and interfaces," Mar. 2017.

\bibitem{bib:ITU-T}\label{Ref:ITU}
ITU-T Rec. G.Sup. 66, ``5G wireless fronthaul requirements in a passive optical network context," Sept. 2020.

\bibitem{bib:Bandyopadhyay}\label{Ref:Bandyopadhyay}
S. Bandyopadhyay \textit{et al.}, ``Review of formation flying and constellation missions using nanosatellites," \emph{Journal of Spacecraft and Rockets}, vol. 53, no. 3, pp. 567-578, Mar. 2016.

\bibitem{bib:Halim}\label{Ref:Halim}
A. U. Chaudhry and H. Yanikomeroglu, ``Free space optics for next-generation satellite networks," \emph{IEEE Consum. Electron. Mag.}, vol. 10, no. 6, Oct. 2021.

\bibitem{bib:Yu}\label{Ref:Yu}
H. Yu and J. Joung, ``Optimization of frame structure and fronthaul compression for uplink {C-RAN} under time-varying channels," \emph{IEEE Trans. Wireless Commun.}, vol. 20, no. 2, pp. 1278-1292, Feb. 2021.

\bibitem{bib:Monserrat}\label{Ref:Monserrat}
J. F. Monserrat \textit{et al.}, ``Multi-radio dual connectivity for 5G small cells interworking," \emph{IEEE Commun. Stand. Mag.}, vol. 4, no. 3, pp. 30-36, Sept. 2020.


\end{thebibliography}
\end{document}